# The Viscosity of Silica Optical Fibres

Li-Yang Shao[1,2], John Canning [1,a], Tao Wang[1,3], Kevin Cook[1] and Hwa-Yaw Tam[2]

[1]*Interdisciplinary Photonics Laboratories (iPL), School of Chemistry, The University of Sydney, NSW, 2006 Australia*

[2]*Department of Electrical Engineering, The Hong Kong Polytechnic University, Hung Hom, Kowloon, Hong Kong*

[3]*Institute of Optoelectronic Technology, Beijing Jiaotong University, Beijing 100044, China*

**Abstract:** The viscosity of an optical fibre over 1000 to 1150 °C is studied by inscribing an optical fibre Bragg grating that can withstand temperatures up to 1200 °C and monitoring fibre elongation under load through the Bragg wavelength shift. This optical interrogation offers high accuracy and reliability compared to direct measurements of elongation, particularly at lower temperatures, thus avoiding significant experimental error. An excellent Arrhenius fit is obtained from which an activation energy for viscous flow of $E_a$ = 450 kJ/mol is extracted; addition of an additional temperature dependent pre-exponential does not change this value. This value is less than that idealised by some literature but consistent with other literature. The log plot of viscosity is overall found to be consistent with that reported in the literature for silica measurements on rod and beams, but substantially higher to past work reported for optical fibres. The discrepancy from an idealised activation energy $E_a \sim 700$ kJ/mol may be explained by noting the higher fictive temperature of the fibre. On the other hand, past optical fibre results obtained by beam bending with much lower values leave questions regarding the method of viscosity measurement and the time taken for structural equilibration. We note that because regenerated gratings already involve post-annealing to stabilise their operation at higher temperature, the structures are much more relaxed compared to normal fibres. This work highlights the need to stabilize components for operation in harsh environments before their application, despite some mechanical compromise. Given the increasing expectation of all-optical waveguide technologies operating above 1000 °C, the need to study the behaviour of glass over the long term brings added significance to the basic understanding of glass in this regime.

---

[a] Electronic mail: john.canning@sydney.edu.au

## I. INTRODUCTION

### A. Fundamental considerations

If there is one singular material that defines the modern age it is silica glass. Silica is the substance whose thermal stability, linked to its very high processing temperature, and optical transparency has enabled, when slightly doped with germanate in most cases, the global village in a way that wireless, for example, could only dream of. There are few that could argue with the de facto recognition of silica, with the 2009 Nobel Prize, as the defining substance of our time.[1] It made optical fibre networks, for communications and sensing, a reality. This is an extraordinary impact for an otherwise random host material, chemically simple, that resists a comprehensive understanding of its formation and all its relaxation processes to this day. It is a material, rapidly formed, whose complexity arises from a heterogeneous distribution of local structures, first illustrated using an extraordinarily inventive and intuitive approach by Bernal involving plasticine balls and an air bladder,[2] with different characteristic relaxations, including local relaxation times in Bernal's analogous local geometries along with the emergent relaxations over longer length scales. This work is a precursor to a series of models, including microcrytallite, quasicrystalline and pseudocrystalline, as alternatives to a Zacchariasan random network.[3] They all exist to explain short range order in glass with long range random packing as revealed, for example, by x-ray scattering studies that show strong overlap between the correlation function of amorphous silica and crystalline silica, particularly cristobalite.[4,5] However, whilst pseudo crystalline labels are convenient, given the short length scales involved (~ 2 nm for cristobalite) this correlation overlap is not strong support for a microcrystalline environment; rather, probably more relevant to local order arising from packing, or tessellation, of similar rigid tetrahedral with similar angles and bond lengths. All of this defies the simplicity and clarity of crystalline organisation over long length scales obtained with much slower relaxation.

The range and distribution of local and emergent processes, both in space and time, makes for a partial theoretical understanding and provides for interesting philosophical questions such as why are local relaxations in the solid state always exponential (it seems regardless of space dimension), retaining many of the characteristics of liquid-like diffusion related to an ambiguous configurational entropy [6] but seemingly over longer time scales. The viscosity is defined in terms of a resistance to flow of a medium and is closely correlated with structural diffusion; activation energy, $E_a$, for flow is often defined. In the strong liquid case limit, it is generally assumed that viscosity, or viscous flow, has a constant activation energy.[6,7] A detailed experimental study of silica viscosity has found this activation differs substantially below and above 1400 °C.[8] Generally, the non-Arrhenius behaviour is associated with the transition between fragile and strong liquids and is believed to be a result of a combination of kinetic and thermodynamic dependencies of structural diffusion, or flow. The particular situation around 1400 °C has been explained by proposing that viscous flow is a result of Si-O line defect motion; below 1400 °C these are related to entropic considerations giving rise to a temperature dependent pre-exponential factor in the Arrhenius equation,[8] whilst above 1400 °C they are unrestricted and constant. No absolute confirmation of this threshold-like behaviour and whether it holds for all thermal histories has been established although the temperature dependent pre-exponential is generally accepted. Why would these assumptions hold, for example, for an optical fibre which has a fictive temperature $T_f \sim$ (1600-1700) °C[9] which lies above 1400 °C, in contrast to most bulk silica glasses where $T_f \sim$ (1200-1300) °C.[10] The optical fibre is rapidly cooled as a consequence of a lager surface area to volume ratio upon drawing and bears closer resemblance to water-quenched glass than those most analyzed.

Overall, the characteristic exponential description of relaxation does allow the total structural relaxation to be well fitted by a stretched exponential function, approximately a sum of single exponentials.[11,12] This distillation of an otherwise complex, and varied, distribution of relaxations has been applied with great success to describe not only bulk glass annealing but also the local annealing of photo-induced change within optical fibre Bragg gratings,[13,14] monitored not by calorimetry but rather by high resolution optical interrogation of the grating spectra as it shifts and decays with glass change. It is indicative of a generic principle of exponential (diffusive) decay and its summation of parts (another contention is where relaxation processes might be expected to intertwine, whether the breakdown is realistic, even if accurate and useful). If the gratings could survive the more general annealing of glass, this latter approach might suggest optical interrogation could be used to characterise "bulk" annealing of the optical fibre itself at temperatures closer to the glass transition, opening up a novel opportunity to explore fibre viscosity and compare this with existing work in the literature. Given the intention of many silica devices set to operate in this regime, this is particularly important.

**B. Practical considerations**

From a practical perspective, the general complexity of glass annealing is what provides the huge range of options available to tailoring glass properties. The rapid drawing and cooling of optical fibres starting above the glass transition temperature, $T_g$ ~1200 °C,[15] leads to compressive stresses at the surface that help mitigate against micro or nano crack propagation, in much the same way toughened glass for cars and bullet proof windows are made. On the other hand, the presence of a softer glass in the core which reduces in volume after having expanded more than the silica cladding in the furnace, usually creates an internal tensile stress at the core-cladding interface – this can be partially overcome by increasing the applied force during drawing so that compression from the outside leads to elongation rather than transverse expansion. Nonetheless, the toughness of optical fibres in resisting surface cracks, further protected from water attack by a polymer coating, is what literally makes the global internet possible today along with a myriad of technologies that continue to seek exploiting this network. Rapid cooling of glass from above $T_g$ is a thermal alternative or complement to chemical toughening of glass, a growing research discipline because of the need for still more robust short-haul high capacity telecommunications and fiber-to-the home (FTTH) technologies, flexible substrates for displays including smartphones and tablets, solar modules and lighting devices, large-sized architectural glazing, lightweight packaging, and more.[16] Despite the promise of new carbon nanotubes technologies with strengths exceeding 100 GPa,[17] vitreous silica with a tensile strength up to 26GPa [18] remains the strongest man-made material that can be produced on a large scale. Optical fibres of pristine surface purity, for example, can reach tensile strengths approaching 14GPa [19] although in practice these values are quickly degraded in telecommunications fibres by surface defects, sometimes through handling alone, to ~ (3-8) GPa.[20] More recently, it has been shown that reduced attenuation in optical fibres can be obtained by high temperature annealing of the fibre.[10,21] On the other hand, others have shown that Rayleigh scattering increases over time in fibres that have experience low temperature thermal aging.[22] All are measures of structural relaxation. However, compromise to the mechanical integrity of the fibre was not reported. The higher temperature regime in Ref.21 is similar to that used to regenerate fibre Bragg gratings [23-27] where mechanical compromise has been established,[28] reducing the original fibre strength from ~5 to ~1.5 GPa – whilst this is acceptable for devices where packaging constraints remain a much more serious issue, for long haul optical fibre communications the disadvantage of up to 70% reduced mechanical integrity outweighs the advantages of reduced signal attenuation. More work is required to optimize the two.

Clearly, the importance of understanding glass remains central to furthering these technological networks underpinned by optical fibre – this is particularly true for the myriad of optical fibre devices that continue to be developed. One of the most significant new areas of research that dramatizes this is the development of practical sensors for harsh (or extreme) environments where optical fibre components need to operate well outside the parameters of telecommunications. In particular, ultra high temperature sensing has attracted considerable attention with the development of new thermally resistant optical fibre components that can withstand temperature above a thousand degrees. In fact, the exploitation of the myriad of relaxations within glass is central to glass-smithing with sub-micron resolution [25,29] where local changes in glass structure and stresses, particularly at the core-cladding boundary within an optical fibre, can be introduced by laser patterning and then exploited through bulk thermal annealing in the presence of a gas, either hydrogen [23-27] or helium.[26] This is the basis of regeneration, or regenerated gratings in optical fibre that can perform in excess of 1200 °C,[24] enabling high temperature optical sensing that can be readily integrated into the coming generation of truly "Smartgrids", linked by the evolving internet. More immediately, the creation of these components led to extraordinary measurements at elevated temperatures, demonstrating their potential to open up new areas and applications. Examples include the first distributed mapping of the temperature within a modified chemical vapor deposition (MCVD) tube under typical optical fibre fabrication conditions where it was discovered that one of the most basic assumptions about heat transfer through the tube wall appears incorrect for typical processing conditions[30] and the measurement of temperature in the engines of heavy duty diesel locomotives used to carry resource loads in Brazil to provide real time feedback to a train driver to prevent the engine from overheating,[31] saving huge costs involved with replacing an engine in the middle of nowhere that has overheated. More broadly, harsh environmental sensing using optical fibre technologies is a rapidly growing field of research and the ability to undertake high temperature diagnosis is a critical element to its success. But all these technologies are operating in a regime close to the fibre transition temperature, where glass viscous flow can take place under load and the material is effectively being annealed. It is therefore essential to understand the viscosity of an optical fibre in this regime particularly with load. On the other hand, such new technologies also offer a novel approach to studying and understanding some of the basic properties of silicate glass directly within optical fibres and waveguide components.

## II. EXPERIMENTAL METHODOLOGY

In this work, a regenerated grating which can operate >1200 ºC for at least several hours is used to study the optical fibre annealing process at elevated temperatures, below the glass transition temperature, $T_g$, in a regime where the relaxation processes coincide with viscosity changes. Regeneration already involves an annealing process close to 900 ºC and, as mentioned earlier, this is known to affect the mechanical strength of the fibre,[28] an indication of a trend towards equilibration of internal strain. All other high temperature fibre component technologies, including femtosecond laser written gratings,[32] will initially suffer the same process during operation so the criticism applied to regenerated gratings is ultimately similar for all fibre components operating in high temperature regimes. Regenerated gratings are in fact pre-equilibrated by additional post-processing at ~1100 ºC to enhance their high temperature performance. Despite this additional higher temperature annealing process, the mechanical integrity is within error unchanged from that of the regeneration process, another indication that some structural relaxation and equilibration of the fibre, with gas, is achieved by the regeneration process alone. The impact on fibre performance in this regime clearly needs to be resolved.

Conventionally, the method for determining various viscosity parameters such as the strain temperature usually involves elongation of the glass sample as a function of load and temperature where the sample is heated uniformly over its entirety.[33, 34] There exists a general standard, ASTM-C336, which is usually followed.[35] Two other approaches use bending or rotation to determine the viscosity of a glass slab or optical fiber by placing either a linear or torsional stress and observing how the material responds. For beam bending this is done by measuring the deflection of the free end of the slab, rod or fiber [8, 21, 36, 37] and for rotation the rate at which twist recovers.[38] In general the determination of the activation energy for viscosity, $E_a$, reported in the literature have tended to agree at least within a factor of three, with the discrepancy highlighting the difficulty in controlling experiments both in terms of keeping contamination out and in different glasses having different thermal histories, or fictive temperatures, and containing different levels of impurities particularly OH. Based on these agreements, Doremus[8] argues compellingly for $E_a$ to be 712 kJ·mol$^{-1}$ below 1400 °C, that obtained by Hetherington *et al.* using dilatation.[39]

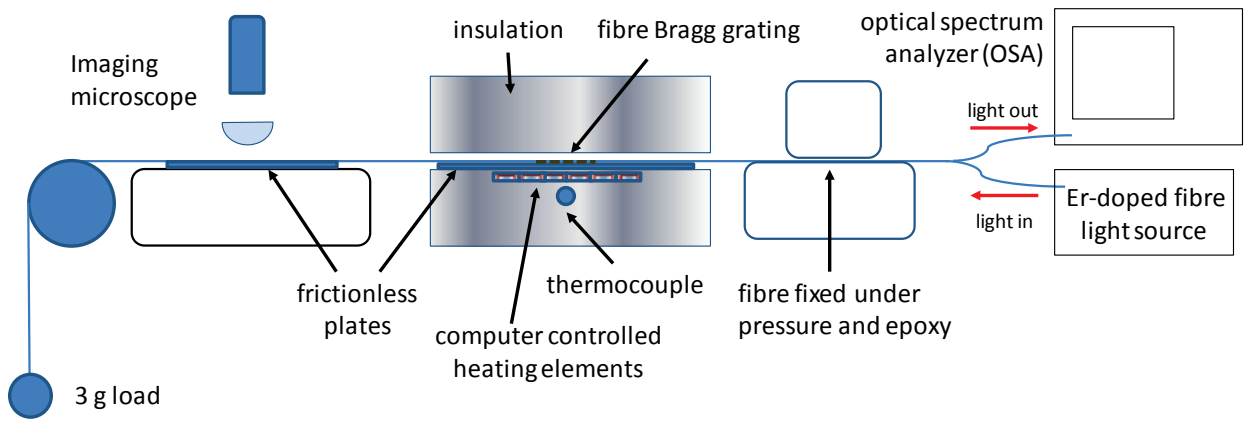

Fig. 1. Schematic of experimental setup used to measure Bragg wavelength shift as a function of load.

Here, we use physical elongation (dilatation) of the fibre under load measured by a CCD microscope, illustrated in Figure 1. The viscosity above 1000 °C can be derived directly by measurements in the rate of change of the regenerated grating spectrum – an optical interrogation method using a broadband source and an optical spectrum analyzer (OSA - a configured spectrometer) allows straightforward determination of a linear rate of change. Although the grating is much shorter than the heated length, it is the change in grating pitch, $\Delta\Lambda/\Lambda$ which is related to the wavelength through the Bragg equation, that concerns us and not the change in length $\Delta L/L$ (explained later). This method is potentially a high precision optical technique for determining and characterising viscosity. It is only possible because the optical grating, a regenerated grating, being used can withstand these high temperatures and the periodic features meet the general criteria that they have a pitch length much larger than the apparent local order in structure seen by the spatial correlation function < 2 nm for both neutron[40] and x-ray scatter.[4,5,41] The grating spectrum therefore will respond unambiguously to the viscous changes within the glass. In the fibre, this glass is a composite system with complications of differing thermal expansion coefficients and local viscosities but it is dominated by an outer cladding with surface compression. The method provides direct and important insight into the way these optical fibres, and therefore silicate glass, can and will perform within environments of unprecedented harshness, both industrial and natural. Needless to say, that whilst the technique is used here to directly evaluate the optical fibre glass itself, the tool can be applied externally to study any manner of materials and systems. It is notable that the fabrication of optical fibres represents not only an important aspect of glass science because it has enabled the global village, but also because it occupies an interesting processing regime where the silica fiber is rapidly quenched whilst being drawn ($T_f \sim 1600$ K) allowing for an unusual amorphous situation with considerably more directional impost than common glass samples that have been analyzed. It is thus of great interest to compare the temperatures at which defined viscous coefficients, such as the annealing and strain points, occur with those in conventional bulk form.

## III. DEFINING PRACTICAL GLASSY PARAMETERS

Despite the continuum of easily altered relaxation processes within glass (with the exception of some cases where relaxation may proceed discretely, such as in high pressure polyamorphic transitions),[42] there is immense value in defining the regimes over which behaviour is viscoelastic or otherwise, both from a technological and a fundamental perspective. Technically, the corresponding annealing and strain points have been practically defined as those at which the viscosity reduces to $\eta = 10^{12}$ and $\eta = 10^{13.5}$ Pa·s respectively.[43] These lie below the general glass transition temperature of silica – ($T_g \sim 1473$ K bulk; $T_f > 1800$ K) where $\eta < 10^{12}$ for cooling rates > 10 K/min. These cooling rates demands a sufficiently long macroscopic measurement timescale ($10^2$ -$10^3$) s from Maxwell's equation.[43] In terms of elongation methods used to characterize optical fibres, for the annealing point these correspond to elongation rates of ~ 0.14 mm/min for a large diameter fibre of ~ 650 μm - such parameters are affected by the cross-sectional area of the fibre. The strain point temperature, $T_s$, is cooler than the annealing point temperature, $T_a$, and so the rate drops dramatically and is roughly 0.032 times that of the annealing point. In general, at the annealing point stresses relax within minutes whilst it takes hours at the strain point. Once cooled slowly below the strain point to avoid fracturing, the glass can be cooled quickly to room temperature – this is approximately the method that has been used to post-stabilize regenerated gratings so that they can operate above 1200 °C.[24] Other important parameters of note are the deformation point, where glass viscosity drops to $\eta = 10^{10.3}$ Pa·s and the glass can be deformed readily, and the softening point, where the glass viscosity drops to $\eta = 10^{6.6}$ Pa·s.

These parameters are all inherently variable with temperature because under different tensions or pressures, glass composition, or fictive temperature,[10] the viscosity is affected and therefore they will differ. In composite systems such as an optical fibre, the problem is further complicated since at the annealing temperature of the silica cladding (for fused silica $T_a \sim 1215$ °C),[43] the core glass may have reached its own glass transition or the softening point, depending on the amount and type of dopants used. Numerous experiments involving regeneration with different fibre types suggests that regeneration probably occurs in the silica cladding mostly with little contribution from the core dopants (beyond the initial laser seed writing phase).[44] In these experiments, standard telecommunications fibre SMF-28 is used – this fibre contains [GeO$_2$] ~ 3 mol %. It is unlikely that the core will affect the results to any appreciable level given that the silica cladding makes up >99% of the fibre, and an outer cladding > 80 %, and will dominate viscosity when the fibre is under tension.

## IV. REGENERATED GRATINGS

Regenerated gratings are a new type of grating with very high thermal stability [23-27]. Descriptively, a relatively low temperature resistant seed grating is produced within the core of an optical fibre by conventional holographic laser inscription, through single photon, two photon or multiphoton processe,[45] and then annealed at an appropriate temperature after an initial ramp. During the annealing process when a gas is present, the seed grating disappears and a new second grating appears - it is this "regenerated" grating which can resist ultrahigh temperatures. The seed grating carries all the information the regenerated grating will carry (assuming uniform annealing) and it is usually a conventional so-called "type I" grating most often written by UV lasers. However, femtosecond laser written gratings have also been regenerated and their stability is also improved.[46] The physical role of the gas, usually hydrogen or helium, is to reduce the existing tensile stress in the fibre during regeneration.[26,29] Hydrogen has the added benefit of significantly enhancing the index modulation of the seed grating - the regenerated grating strength directly correlates with the seed grating strength. To further stabilise the regenerated gratings and improve their temperature performance, post-annealing above the strain temperature of silica ($T_s \sim 1070$ °C) is undertaken.[24,29,47] In our previous work, it was shown that gentle tension can lead to large wavelength spread between the regenerated grating and post-annealing[27] so the potential for viscous studies is clear. In this work, we examine the strain and derive viscosity information of the fibre as a function of temperature over the range 1000 °C to 1150°C, < $T_g$. This spans that regime where the glass stresses can relax; without load there is no elongation or deformation.

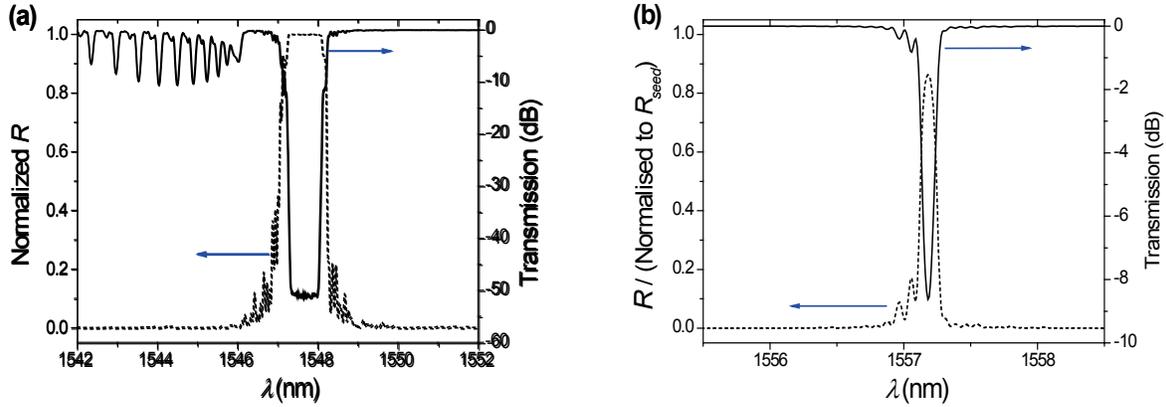

Fig. 2. Typical reflection (dashed) and transmission spectra of (a) seed grating at 20°C and (b) regenerated grating at 850°C.

Strong seed gratings were inscribed directly through a phase mask into SMF-28 fiber preloaded with $H_2$ ($P$ = 180 atm, $T$ = 80 °C, $t$ = 4 days) using 193 nm from an ArF laser ($E_{pulse}$ = 67 mJ/cm$^2$; $f_{cum}$ = 241 J/cm$^2$ ; $RR$ = 30 Hz; $\tau_w$ = 15 ns). Details of the procedures can be found in various references.[23-29] The optical spectra of the gratings were measured using a broadband light source and an optical spectrum analyzer (OSA, resolution: 0.05nm) shown in Fig. 1. Fig. 2(a) shows a typical seed reflection and transmission spectra in the near infrared for all cases studied here.

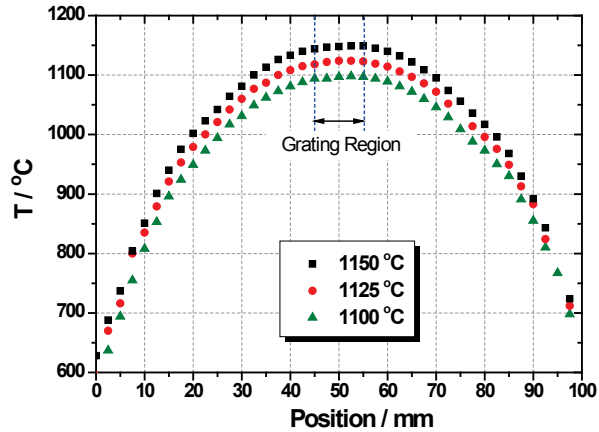

Fig. 3. Temperature distribution of the hot zone in the heater (three representative temperatures, 1100 °C, 1125 °C, 1150°C). The marked region shows where the grating located during the annealing process.

For regeneration, the seed gratings were annealed within a customised annealing oven (Fig. 1.). Fig. 3 shows the temperature distribution of the hot zone in the heater and the relative position of the gratings - there is a weak quadratic variation of temperature over the grating length. The profile in the centre is approximately the same for all temperatures but scaling with each 25 ºC increment. To compare the effect of strain on the regeneration, one seed grating has minimal tension applied to ensure the fibre is straight during the heating of gratings while another had a load of 3 grams (g) applied to it. The annealing schedule for regenerating the seed grating is shown in Fig 4(a), right axis, where the temperature of the furnace was raised uniformly from room temperature to ~850°C (the erasing temperature for the seed grating during regeneration) in one hour and kept constant at that temperature for 180 minutes. Fig. 2(b) shows the typical spectra of regenerated grating after the whole regeneration process at 850°C. When the regeneration process is saturated, the furnace was heated up to 1100 °C in $t = 20$ min and kept constant for 160 minutes during the subsequent annealing phase. Fig. 4(a) & (b) show the evolution of the peak reflection strength, $R$ (normalized to the maximum strength), and Bragg wavelength shift, $\Delta\lambda_B$, of the fibre grating during regeneration and annealing with and without 3g load.

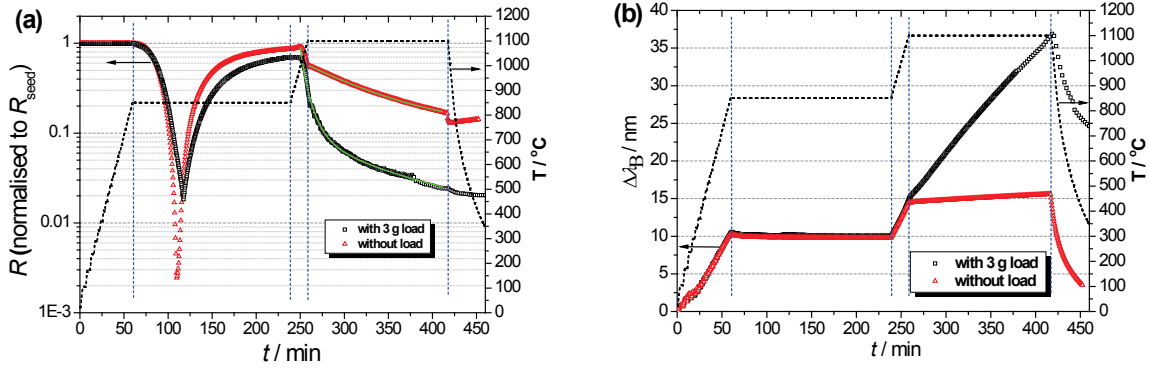

Fig. 4. Evolution of (a) peak reflection and (b) Bragg wavelength shift of one grating during regeneration process (post-annealed temperature: 1100 °C; black square: with 3 gram load; pink triangle: without load).

From Fig. 4(a), the regeneration rollover threshold appears shifted to longer times and $R$ slightly reduced for the fibre with the load. Much more significant differences become obvious when further annealing at 1100 °C, which is kept constant, where all decays are easily fitted by single exponentials. The rate of decay at this temperature is much faster for the case with load. Given the decay is single exponential (Fig. 4(a)) we can correlate the associated single relaxation directly with the observed large wavelength shift, $\Delta\lambda_B$, shown in Fig. 4(b). For 160 min of post-annealing, $\lambda_B$ has shifted over 21 nm under the 3 g load, while it has only shifted 1 nm with minimal tension used to keep the fibre straight. $\Delta\lambda_B$ is linear with time and the rate of change is calculated to be $d\lambda_B/dt \sim \Delta\lambda_B/t \sim 0.134$ nm/min. To explain these results requires elongation to be present under load; the rate of change and the corresponding wavelength shift is therefore a direct measure of the local viscosity changes experienced by the grating. This was confirmed experimentally by measuring elongation using a micrometer and a small portable CCD microscope (see section V).

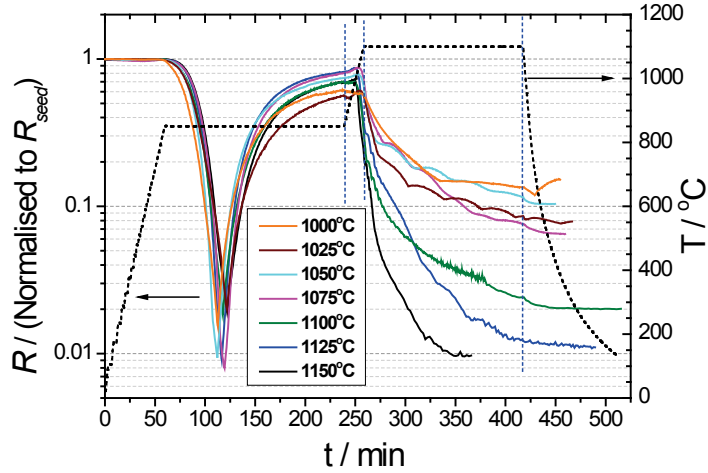

Fig. 5 Evolution of peak reflection, *R*, of gratings during regeneration process with different post-annealing temperature from 1000 °C to 1150 °C under 3 gram load. *R* is normalized to the seed grating reflectivity, $R_{seed}$

To obtain information as a function of temperature, these experiments were then repeated under similar experimental conditions but changing the post processing temperature in intervals of 25 °C over 1000 to 1150 °C. Fig. 5 shows the evolution of the peak reflection of the SMF28 gratings during regeneration at different post-annealing temperatures under a 3 g load. The characteristic curves in Fig. 5 show the trend where the rollover threshold for regeneration is weaker, the regeneration reflection peak decreases and the decay with post-annealing accelerates. Each of the decay curves can again be fitted by a single exponential where the only variable is the rate of decay. This is consistent with a single relaxation process for each case and is also consistent with softening of the glass under a fixed load. The corresponding elongations are summarised in Table I.

## V. VISCOSITY FROM REGENERATED GRATING CHARACTERISTIC CURVES

Fig. 6(a) summaries the evolution of $\Delta\lambda_B$ during regeneration for all the post-annealed temperatures spanning 1000 °C to 1150 °C under a 3 g load. For the gratings post-annealed at 1125 °C and 1150 °C, the annealing time was shorter since $\Delta\lambda_B$ went outside the bandwidth of the broadband source used to interrogate the gratings and therefore could not be measured. Interestingly, by extrapolation the change at 1150 °C will be > 100 nm which offers, for example, an extraordinarily powerful method of tuning the grating wavelength almost anywhere across the telecommunications window.

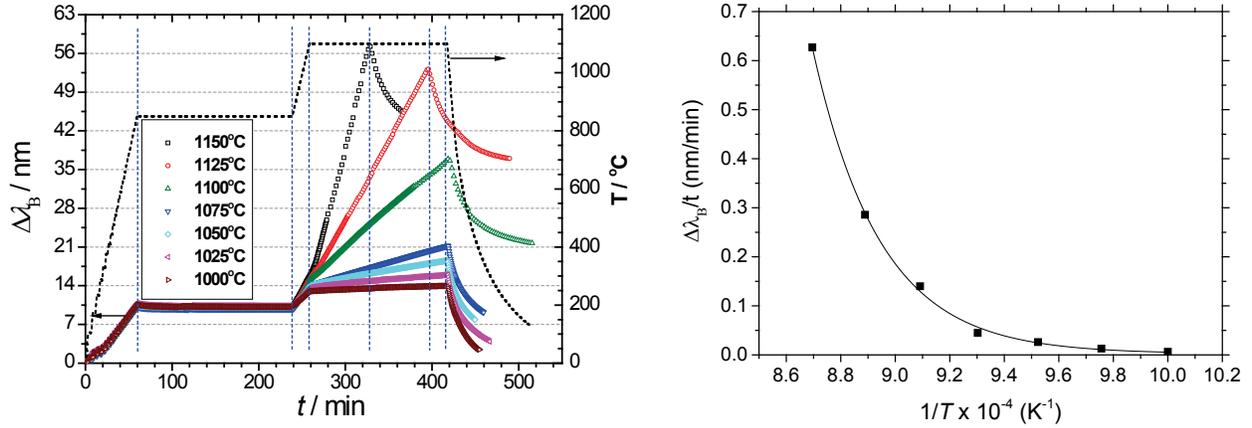

Fig. 6. (a) Evolution of the Bragg wavelength shift of the gratings during regeneration with different post-annealing temperatures from 1000 to 1150 °C under a 3 g load; (b) Rate of change with time of $\Delta\lambda_B$ increases as a function of post-annealing temperature.

Table I summarises the data for both the cases annealed with a 3 g load and without any load other than applied tension to straighten the fibres. The rate of change as a function of inverse temperature is shown in Fig. 6(b). All the response profiles during the period of post-annealing are linear with constant temperature.

TABLE I. Summary of Bragg wavelength shifts with and without load (3 g) at different temperatures.

| T /°C | 1000 | 1025 | 1050 | 1075 | 1100 | 1125 | 1150 |
|---|---|---|---|---|---|---|---|
| $\Delta\lambda_B$ (3 g) /nm | 1.02 | 2.06 | 4.19 | 7.19 | 21.45 | 37.13 | 40.04 |
| $\Delta\lambda_B$ (0 g) /nm | 0.31 | 0.43 | 0.63 | 0.86 | 1.05 | 2.4 | 2.1 |

The viscosity, $\eta$, can be defined in terms of a strain rate, $d\varepsilon/dt = (1/L_0) \cdot dL/dt$, corresponding to an applied stress, $S = F/A = ma/A$ where m is the mass (3 g), $a$ is the acceleration (gravity g = 9.8 m·s$^{-2}$) and $A$ is the cross-sectional area of the fibre = $1.227*10^{-8}$ m$^2$. Given very low volume expansion coefficient of silica, the stress can therefore be calculated to be $S = 2.4*10^6$ kg·m$^{-1}$·s$^{-2}$ (MPa). The optical fibre pulled by $S$ (the 3 g load) elongates at a rate $dL/dt$, where $L$ is the length of fibre being stretched and $t$ is the time over which that occurs. The viscosity is:

$$\eta = \frac{SL_0}{dL/dt} = \frac{S}{d\varepsilon/dt} \tag{1}$$

The elongation data measured by CCD microscope for 1100, 1125 and 1150 °C is summarized in Fig. 7 – it is derived from the average of three measurements at each temperature. In practice it was very difficult to measure with great accuracy physical elongation below 1100 °C over short intervals indicating some of the potential error in deriving viscosity from physical displacement measurements alone.

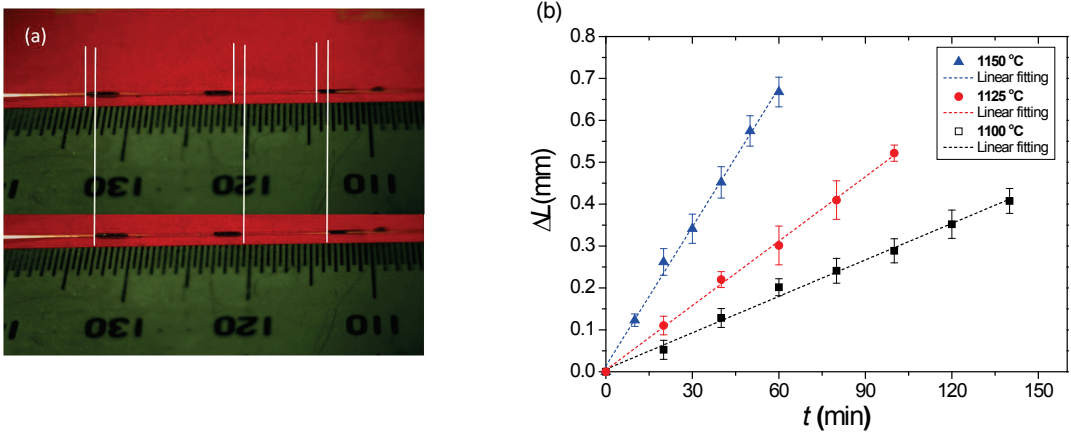

Fig. 7. Measurement of the physical elongation of the fibre: (a) 3 points along the fibre are marked and monitored by CCD microscope during annealing. The elongation is taken as the average of the three – this is the image for 1423 K; (b) Plot of rate of elongation over time at three temperatures. Below 1373 K, the measurement resolution was insufficient to be worthwhile.

Alternatively, from the Bragg equation $\Delta\lambda_B = 2n_{\text{eff}}\Delta\Lambda$, it would seem straightforward to see that:

$$\frac{\Delta\lambda_B}{\lambda_B} = \frac{\Delta\Lambda}{\Lambda} = \frac{\Delta L}{L} \tag{2}$$

This equation shows there will be a corresponding change in Bragg wavelength associated with a stretch of the fibre. However, as well as physical elongation, the measured Bragg wavelength shift will also see a change in the local refractive index such that the Bragg equation is more accurately described as:

$$\Delta\lambda_B = 2n_{eff}(T,\varepsilon) \cdot \Delta\Lambda \tag{3}$$

The effective refractive index is therefore also affected by the applied load and the $T$ the grating is subjected to. Specifically, the relative change in Bragg wavelength, $\Delta\lambda_B/\lambda_B$, is affected both by the elasto-optic coefficient of silica, $p_e$, ≈ 0.22, and the thermo-optic coefficient, $\kappa \approx 5 \times 10^{-6}$. For the changes in wavelength, the initial Bragg wavelength was taken at the start of the elevated constant temperature regime where the rate of spectral change was measured and from which viscosity can be calculated; any thermo-optic contribution is removed from the measured shifts. Hence, the observed relative change of the wavelength at constant temperature is therefore dependent only on the strain:

$$\frac{\Delta \lambda_B}{\lambda_B} = (1 - p_e)\varepsilon \tag{4}$$

By localizing the relative change to the grating period, the advantage of this method is that it avoids the need to calibrate the length of heated region as a function of hot zone distribution at each temperature. Thus the viscosity in this constant temperature window can be described in terms of the optical wavelength shift with stress as:

$$\eta = \frac{S(1-p_e)\lambda_B}{d\lambda/dt} = \frac{0.78 S \lambda_B \Delta t}{\Delta \lambda_B} \tag{5}$$

At a given temperature Fig. 6 shows that there is a constant rate of change in wavelength, $\Delta\lambda_B/\Delta t$, for a constant applied stress, $S$, and therefore a particular viscosity, $\eta$. The calculated viscosity obtained from the experimental data is plotted in Fig. 8(a). A reasonable Arrhenius fit is obtained where $A$ is a fitted pre-exponential scaling factor and $E_a$ is the activation energy for viscous flow, which has some relation to the configuration entropy in the Adams-Gibbs reformulation: [11,12,47]

$$\eta = A e^{\left(\frac{E_a}{RT}\right)} \tag{6}$$

However, the general form which can describe accurately over the entire temperature span involves a temperature dependent pre-exponential term, so the scaling factor is also affected by the activation energy and temperature: [8, 43]

$$\eta = AT e^{\left(\frac{E_{a1}}{RT}\right)} \left[1 + B e^{\left(\frac{E_{a2}}{RT}\right)}\right] \tag{7}$$

We note that the original Doremus' expression[8] assumes no $T$ scaling in the above expression since the experimental error is often larger than the small differences with and without this factor. This expression has also been recast in terms of the configurational entropy.[49] Both Fig. 8(a) and (b), which shows the log of the viscosity in Pa·s, suggest the additional pre-exponential temperature dependence is unnecessary in this temperature regime. The activation energy from the standard single exponential Arrhenius fit is found to be $E_a$ = 450 kJ.mol$^{-1}$ whilst for the addition of a secondary relaxation process in the pre-exponential term $E_a = E_{a1} + E_{a2} = 459$ kJ·mol$^{-1}$ with 1/T scaling and 450 without 1/T (Doremus) scaling slightly reduced. The small difference falls within experimental variation and so it can be assumed the viscosity follows the Arrhenius expression well in this strong glass regime. These values are lower by about 10 to 40 % than that reported by other authors for bulk glass, summarised by Doremus[8] who favours the higher value of Hetherington *et al.*[39] where $E_a$ = 712 kJ/mol over 1100 – 1400 °C, and lower values $E_a \sim 500$ kJ/mol above 1400 °C. Compared to other measurements in silica optical fibres, we note our viscosity is an order of magnitude higher than those reported in Ref. 21 for example.

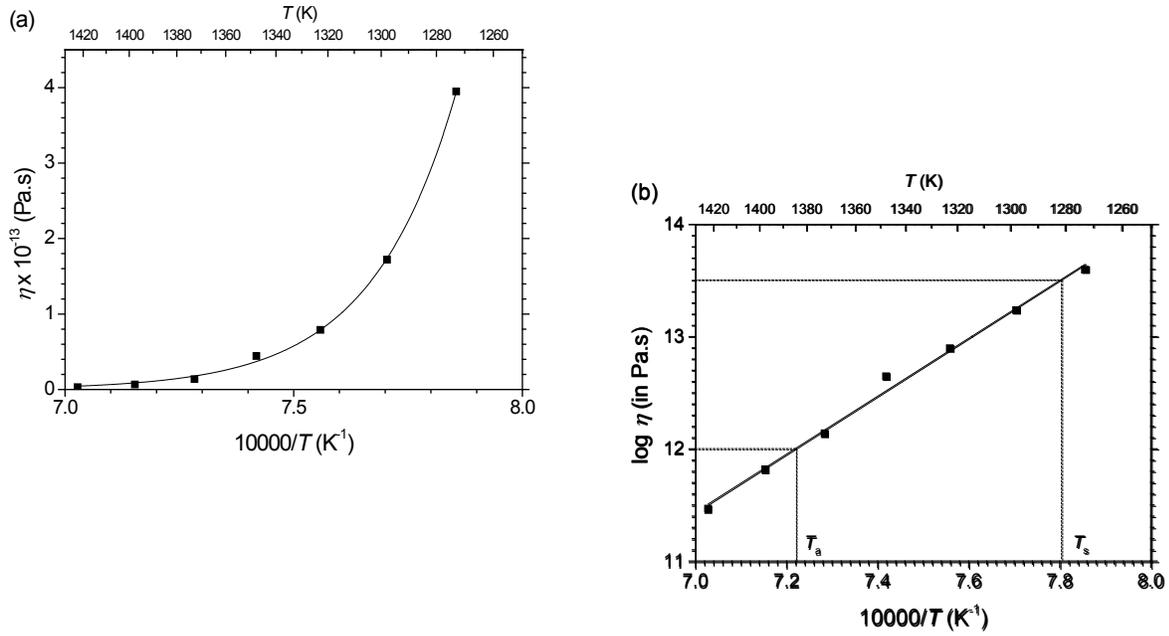

Fig. 8. Viscosity curves for the optical fibre with regenerated grating: (a) an excellent Arrhenius fit is obtained; (b) the log plot of the viscosity in Pa·s. The annealing and strain temperatures are determined to be $T_a \sim 1114$ °C and $T_s \sim 1010$ °C.

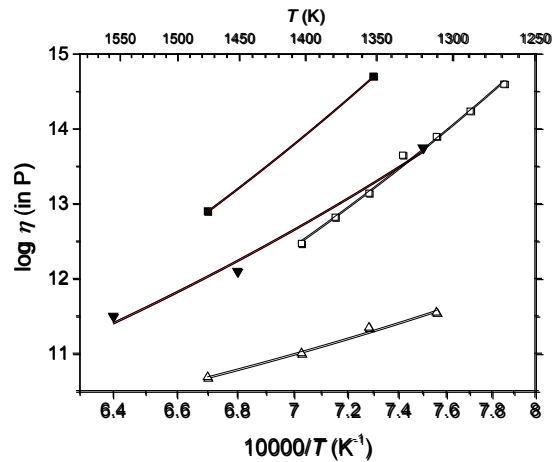

Fig. 9. Log viscosity curves in Poise for work reported here and reproduced for previous work: ■ From Hetherington et al. [37]; □ This work; ▼ From Urbain et al. [51] – data point at 7.5 is a linear extrapolation of their data at higher T; Δ From Sakaguchi & Tadori [19]. The data from this work is taken for 2-10 min thermal exposure, seemingly too short to allow equilibration of glass structure.

## VI. DISCUSSION

The good Arrhenius fit of the viscosity versus inverse temperature curve, essentially the traditional VFT equation, where no secondary scaling of the pre-exponential term *A* with temperature is required, is consistent with silica as a so-called "strong" liquid[50] in this temperature regime. Despite the differences in the thermal history of optical fibre fabrication relative to other larger volume silica, the general thesis that the Arrhenius fit holds above 1000 °C at least until the glass transition temperature[51] appears valid. The calculated value of the activation energy for viscous flow, is $E_a \sim 450$ kJ/mol – this is substantially less than Doremus' preferred value of $\sim 712$ kJ/mol[7] as calculated by Hetherington *et al.* for bulk silica.[38] The viscosity measurements over the temperature range measured are a little lower than the data reported by Bruckner *et al.*[37] and others for silica.[51, 52] The viscosity profiles shown in Fig. 9 allow a comparison between this work and that previously reported. Doremus[8] argues that a high fictive temperature and insufficient relaxation is probably to blame for lower figures of results from that of Hetherington *et al.*[39] who measured $\sim 712$ kJ/mol. Groups such as Urbain *et al.*[53] report $E_a \sim 551$ kJ/mol from 1200 to 2400 °C and Fontana & Plummer[54] report a similar value $E_a \sim 519$ kJ/mol using both their own data and that of Bruckner.[55] In our case we do not believe this argument is applicable. In the first instance, the higher fictive temperatures of optical fibre fabrication might suggest lower viscosities than expected in the range we are examining – Sakaguchi & Tadori[21] take this argument for granted to explain their substantially lower values. Their data, however, appears to have been taken over relatively short timescales (2-30 min) with no pre-equilibration stage. In our results we have demonstrated a constant rate of change in viscosity so we are confident about structural equilibration. Further, there is an annealing phase during regenerated grating production, followed by a subsequent post-annealing phase which is used to stabilise regenerated gratings for high temperature operation. It is almost certain that our glass is sufficiently relaxed by comparison. Such relaxation leads to weakening of the fibre since compressive stresses are annealed out. The regenerated grating results have shown that even annealing ~800 – 950 ºC to obtain regeneration weakens the mechanical strength of the fibre from ~ (5-6) GPa to ~ (1.5-2) GPa, a 30% reduction. The reasons for this reduction are not thought to be related to water contamination (OH is well known to reduce viscosity curves):[55]

regeneration was performed in a dry atmosphere and as well a reduced strength was obtained for He regeneration as for $H_2$.

Implicit in a relaxation argument is that simple viscosity measurements done in too short a timescale will not be accurate precisely because the optical fibre glass is not relaxed at these temperatures and viscosity is underestimated. This in turn raises fundamental questions about the suitability of optical fibres for long term operation in harsh environments above their typical specifications for telecommunications (<200 ºC). As noted earlier, this is fine for specialized components where packaging remains a much more serious issue over the fibre itself, including sensors and lasers operating in harsh environments. On the other hand long lengths of optical fibre within networks designed to operate in harsh environments will experience changes over time – along with hydrogen penetration, we predict this may in part be a key contributor to the observed degradation of optical fibres down oil bores, for example.[56] Consequently, some of the criticism of regenerated gratings compared to high temperature gratings produced by other means, such as with femtosecond laser writing, are redundant given that regeneration involves in advance a natural relaxation in the glass that will accompany all optical fibre and waveguide devices operating in harsh environments (at least at temperatures as low as the regeneration temperature – which can be less than 800 ºC). Long term annealing and testing is clearly still necessary for the development of silica based harsh environment sensors and devices.

The impact of frozen-in strain along the fibre within the hot zone profile may contribute to the slightly lower viscosity we measure. It has been observed with increasing tension, both linear and twist, that there is an increasing anomalous fibre contraction during elongation response beginning around 700 °C within some fibres,[57,58] reaching a peak ~ 1000 °C. This contraction is consistent with glass polymerization analogous to long range networking found in polymers, which are often drawn, and which can also lead to reduced viscosity.[59] Such behaviour will seriously complicate viscosity measurements if the glass is not fully relaxed and can explain the low viscosity reported in Ref. 21. Given the annealing stages of our own SMF-28 fibre gratings, the linear evolution of the Bragg wavelength and the strong Arrhenius fit there is no evidence of this anomaly impacting our data, emphasizing the importance of allowing relaxation.

The effects of frozen in strain are also observed as, for example, phosphosilicate phase separation and polymerization in phosphosilicate fibres[60] where polymerization of the glass is thought to occur. In most silicate optical fibres, normal ring cavity nanopores making up the glass network can become nanotubes aligned along the direction of the fibre so it is difficult to imagine that this does not affect viscosity in a directionally preferential manner. It breaks the isotropic nature assumed behind viscous flow within a random network material and may explain why our values are slightly lower although the rate of change similar, to Hetherington *et al.* data.[39]

If all these factors act as an additional spring on the system the magnitude of the effective load on the grating may be different to the actual applied load and there will be a corresponding change in viscosity. This needs consideration since our elongation method involves localized annealing with extended non-uniform temperature profiles along the fibre, unlike bulk glass measurements where the entire glass is supposedly subjected to uniform heating. We believe the analysis of the grating pitch, and averaging out of other contributions, minimize such a convolution of effects.

The differences in viscosity compared to other work will give rise to differences in annealing and strain points. Using the definitions described earlier, from Fig. 8(b) we can work out the annealing and strain temperatures to be $T_a \sim 1114$ °C and $T_s \sim 1010$ °C. As expected these are a little lower than reported values for fused quartz of $T_a \sim 1140$ °C and $T_s \sim 1070$ °C and fused silica $T_a \sim 1215$ °C and $T_s \sim 1120$ °C.[43,61]

## VII. CONCLUSION

The all optical interrogation of grating structures designed to withstand the temperature window used, combined with a precision small-volume heater, has offered an accurate and reproducible way to chart viscosity with high resolution whilst using a well-established elongation method. This was superior to direct visual measurements of elongation through a CCD microscope. The viscosity values obtained are higher than previous fibre results commensurate with but slightly lower than those obtained for bulk fused quartz. To the best of our knowledge, it is the first time the annealing and strain temperatures for an optical fibre have been derived. The fabrication of regenerated gratings is underpinned by glass relaxation so that the glass is stabilised prior to high temperature use. Our work suggests strongly that without such stabilization, other methods of fabricating sensors and devices for high temperature performance will suffer degradation over time as the fibre relaxes. We therefore recommend pre-annealing of all devices prior to high temperature operation, the amount and procedure for pre-annealing will be sensitive to the temperature and duration of anticipated operation. Finally, the work reported here demonstrates the need for ongoing research in evaluating and understanding glass changes and exploring the opportunity for customizing these changes to suit specific functionality for hash environments. Regenerated gratings are already one such example of this understanding, proving to be an invaluable tool for further exploration of glass science.

**ACKNOWLEDGEMENT**

The project acknowledges Australian Research Council (ARC) FT110100116 grant funding. L. Shao acknowledges the award of an Australia Award Endeavour Research Fellowship, the Hong Kong Polytechnic University project G-YX5C and the National Natural Science Foundation of China under Grant No. 61007050. T. Wang acknowledges the Visiting Scholarship Award from China Scholarship Council (CSC).